\documentclass{elsart5p}

\usepackage{graphicx}

\usepackage{amssymb}
\usepackage{amsmath}

\begin{document}

\begin{frontmatter}
\title{Spin-polarons in an exchange model}

\author[aff1]{O. Navarro}
\author[aff2]{E. Vallejo\corauthref{cor1}}
\corauth[cor1]{Corresponding author.}
\ead{emapion@yahoo.com}
\author[aff3]{M. Avignon}
\address[aff1]{Instituto de Investigaciones en Materiales, Universidad Nacional Aut\'{o}%
noma de M\'{e}xico, Apartado Postal 70-360, 04510 M\'{e}xico D. F., M\'{e}%
xico.}
\address[aff2]{Facultad de Ingenier\'{i}a Mec\'{a}nica y El\'{e}ctrica, Universidad Aut\'{o}noma de Coahuila.
Carretera Torre\'{o}n-Matamoros Km. 7.5 Ciudad Universitaria C.P. 27276 Torre\'{o}n, Coahuila.}
\address[aff3]{Institut N\'{e}el, CNRS and Universit\'{e} Joseph Fourier, Boite Postale 166, 38042 Grenoble, France}

\begin{abstract}
Spin-polarons are obtained using an Ising-like exchange model consisting of double and super-exchange interactions 
in low dimensional systems. At zero temperature, a new phase separation between small magnetic polarons, one conduction electron 
self-trapped in a magnetic domain of two or three sites, and the anti-ferromagnetic phase was previously reported. 
On the other hand the important effect of temperature was missed. Temperature diminishes Boltzmann probability 
allowing excited states in the system. Static magnetic susceptibility and short-range spin-spin correlations at zero magnetic field were calculated
to explore the spin-polaron formation. At high temperature Curie-Weiss behavior is obtained and compared with the
Curie-like behavior observed in the nickelate one-dimensional compound $Y_{2-n}Ca_{n}BaNiO_{5}$. 

\end{abstract}

\end{frontmatter}

\section{Introduction}\label{}

Phase transition in a given physico-chemical system is characterized by 
parameters like the range of the microscopic interactions, the space dimensionality $d$ and the
dimensionality of the order parameter, often referred to the spin dimensionality $s$. There are
features whose qualitative nature is determined by the universality class to which the system belongs.
Short-range interactions, double and super-exchange nearest-neighbor type, classical and quantum spins $s$
in $d$-dimensional systems have been studied \cite{zener1951,anderson1955,degennes1960,yunokidagotto1998,yunoki1998,
yamanaka1998,batista19982000,koshibae1999,garcia20002002,aliaga2001,neuber2006,VLNA2009,SVNA2009}. 
Double-exchange (DE) interaction or indirect exchange, is the
source of a variety of magnetic behavior in transition metal and rare-earth
compounds\cite{Mattis2006}. The origin of DE lies in the
intra-atomic coupling of the spin of itinerant electrons with localized
spins $\overrightarrow{S}_{i}$. This coupling favors a ferromagnetic (F) background of local spins and may lead to 
interesting transport properties such as colossal magnetoresistance. This mechanism has been widely used in the
context of manganites \cite{zener1951,anderson1955,jonker1950}. This F
tendency is expected to be frustrated by anti-ferromagnetic (AF) inter-atomic super-exchange (SE) interactions
between localized spins $\overrightarrow{S}_{i}$ as first discussed by de
Gennes\cite{degennes1960} who conjectured the existence of canted states.\
In spite of recent interesting advances, our knowledge of magnetic ordering
resulting from this competition is still incomplete.

Although it may look academic, the one-dimensional (1D) version of this
model is very illustrative and helpful in building an unifying picture. On
the other hand, the number of pertinent real 1D systems as the nickelate
one-dimensional metal oxide carrier-doped compound $Y_{2-n}Ca_{n}BaNiO_{5}$%
\cite{DiTusaKojima} is increasing. Haldane gap $(\sim 9meV)$ has been observed for the parental compound $n=0$ $Ni^{2+}$ (S=1) from 
susceptibility and neutron scattering measurements. In these compounds, carriers are
essentially constrained to move parallel to $NiO$ chains and a
spin-glass-like behavior was found at very low temperature $T\lesssim 3K$
for typical dopings $n\approx0.04$, $0.1$ and $0.15$. At high temperature Curie-like behavior
of the magnetic susceptibility was found. The question is how physical properties change by introducing $n$ holes in the system. 
In the doped case the itineracy of doped electrons or holes plays an important role taken into
account by the double-exchange mechanism.
Recently, it has been shown
that three-leg ladders in the oxyborate system Fe$_{3}$BO$_{5}$ may provide
evidence for the existence of spin and charge ordering resulting from such a
competition\cite{vallejo2006}.\newline

Naturally, the strength of the magnetic interactions depends significantly
on the conduction electron band filling, $x=1-n$. At low conduction electron density, F
polarons have been found for localized $S=1/2$ quantum spins \cite%
{batista19982000}. \textquotedblleft Island\textquotedblright\ phases,
periodic arrangement of F polarons coupled anti-ferromagnetically, have been
clearly identified at commensurate fillings both for quantum spins in one
dimension \cite{garcia20002002} and for classical spins in one\cite%
{koshibae1999} and two dimensions \cite{aliaga2001}. Phase separation
between hole-undoped antiferromagnetic and hole-rich ferromagnetic domains
has been obtained in the Ferromagnetic Kondo model \cite{yunokidagotto1998}.
Phase separation and small ferromagnetic polarons have been also identified
for localized $S=3/2$ quantum spins \cite{neuber2006}. In addition to the expected F-AF 
phase separation appearing for small super-exchange coupling, a new phase separation between small
polarons ordered (one electron within two or three sites) and AF regions for larger SE coupling was found \cite{VLNA2009,SVNA2009}. These phase
separations are degenerate with phases where the polarons can be ordered or not giving a natural response to the instability
at the Fermi energy and to an infinite compressibility as well. 
Wigner crystallization and spin-glass-like behavior were also obtained and could explain the spin-glass-like behavior observed 
in the nickelate 1D doped compound $Y_{2-n}Ca_{n}BaNiO_{5}$ \cite{VLNA2009}.

In this paper, we present a study of the parallel static magnetic susceptibility in an Ising-like exchange model. 
Short-range spin-spin correlations are also presented. Our results are compared with the Curie-like behavior observed at high 
temperature in the nickelate one-dimensional compound $Y_{2-n}Ca_{n}BaNiO_{5}$ \cite{DiTusaKojima}. 
The paper is organized as follows. In section II a brief
description of the model is given. In section III, results and a discussion
are presented. Finally, our results are summarized in section IV.

\section{The model}

The DE\ Hamiltonian is originally of the form, 
\begin{equation}
H=-\sum_{i,j;\sigma }t_{ij}(c_{i\sigma }^{+}c_{j\sigma }+h.c.)-J_{H}\sum_{i}%
\overset{\rightarrow }{S_{i}}\cdot \overset{\rightarrow }{\sigma }_{i},
\label{A}
\end{equation}%
where $c_{i\sigma }^{+}(c_{i\sigma })$ are the fermions creation
(annihilation) operators of the conduction electrons at site $i$, $t_{ij}$
is the hopping parameter and $\overrightarrow{\sigma }_{i}$ is the
electronic conduction band spin operator. In the second term, $J_{H}$ is the
Hund's exchange coupling. Here, Hund's exchange coupling is an intra-atomic
exchange coupling between the spins of conduction electrons $\overrightarrow{%
\sigma }_{i}$ and the spin of localized electrons $\overrightarrow{S}_{i}$. 
This Hamiltonian simplifies in the strong coupling limit $J_{H}\rightarrow
\infty $, a limit commonly called itself the DE model. In this strong coupling 
limit itinerant electrons are now either parallel or anti-parallel to local spins and are thus spinless. The
complete one dimensional DE+SE Hamiltonian becomes, 
\begin{equation}
H=-t\sum_{i}(\cos \left( \frac{\phi _{i,i+1}}{2}\right)
c_{i}^{+}c_{i+1}+h.c.)+J\sum_{i}\overset{\rightarrow }{S_{i}}\cdot \overset{%
\rightarrow }{S_{i+1}},  \label{C}
\end{equation}
$\phi _{i,i+1}$ is the relative angle between localized spins at
sites $i$, $i+1$ defined with respect to a z-axis 
taken as the spin quantization axis of the itinerant electrons. The super-exchange coupling is an anti-ferromagnetic inter-atomic
exchange coupling between localized spins $\overrightarrow{S}_{i}$. This coupling is given in the second term of the former equation. 
Here $J$ is the super-exchange interaction energy. An Ising-like model with itinerant electrons will be considered in this paper, i. e. 
$d=1; s=1$  and $\phi_{i}=0$ or $\pi$. For itinerant electrons (holes) an electron (hole)-single approximation will be used.
The nickelate one-dimensional parental compound $Y_{2}BaNiO_{5}$, is basically formed of quasi one-dimensional chains of $Ni^{2+}$. 
$3d_{3z^{2}-r^{2}}$ and $3d_{x^{2}-y^{2}}$ are two relevant $Ni^{2+}$ orbitals in this system. $3d_{x^{2}-y^{2}}$ is basically localized while
$3d_{3z^{2}-r^{2}}$ has finite overlap with $2p_{z}$ orbital of the O \cite{M1993P1995}. So, to make contact with 
the nickelate one-dimensional compound $Y_{2-n}Ca_{n}BaNiO_{5}$, $N$ localized S=1/2 spins in the $3d_{x^{2}-y^{2}}$ orbital will be considered. 
On the other hand itinerant electrons $x$ or holes $n$ will be placed in the $3d_{3z^{2}-r^{2}}$ orbital. The role of these electrons (holes) within
the parental compound $n=0$, will be considered by the DE mechanism. Within our Ising-like model there is an electron-hole symmetry.
 
Exact parallel static magnetic susceptibility $\chi$ and short-range spin-spin correlations are presented using a standard canonical ensemble. 
To obtain $\chi$ within the electron (hole)-single approximation is necessary to calculate eigenvalues of the following matrix

\begin{equation}
\mathbf{H}=
\left( \begin{array}{ccccc}
h_{1} & t_{1,2} & 0 & 0 & \ldots\\
t_{2,1} & h_{2} & t_{2,3} & 0 &\ldots\\
0 & t_{3,2} & h_{3} & t_{3,4} &\ldots\\
0 & 0 & t_{4,3}& h_{4}&\ldots\\
\vdots& \vdots& \vdots& \vdots& \ddots     
\end{array}\right)  \label{Hamil}
\end{equation}

where 

\begin{equation}
h_{i}=JS^{2}\sum_{k=1}^{N-1}\cos \left( \phi_{k}-\phi_{k+1} \right)-\mu B \sum_{k=1}^{N}\cos \left( \phi_{k}\right)-\mu B \cos \left( \phi_{i}\right),
\end{equation}

in the former equation first term is super-exchange interaction and the second one is the Zeeman coupling of the localized background of S=1/2
spins. Third term is the coupling between the magnetic moment $\mu$ of the itinerant electron and the magnetic field $B$.
A magnetic field was introduced to calculate $\chi$.

\begin{equation}
t_{i,j}=t_{j,i}=-t\cos \left( (\phi_{k}-\phi_{k+1})/2 \right).
\end{equation}

With eigenvalues of equation (\ref{Hamil}) is easy to obtain partition function $Z$ in the canonical ensemble within the electron-single approximation 
\begin{equation}
Z=\sum_{i<j<k,\cdots} e^{-\beta (\in_{i}+\in_{j}+\in_{k}+\cdots )}.
\end{equation}
For one ($i$), two ($i$ and $j$), three  ($i$, $j$ and $k$) and ($\cdots$) itinerant electrons respectively.  $\beta =\frac{1}{k_{B}T}$ being $k_{B}$ Boltzmann constant and $T$ temperature$.$

Magnetic susceptibility is related with partition function as

\begin{equation}
\chi=Limit_{B \rightarrow 0} \hspace{0.2cm} k_{B}T \frac{\partial^{2}}{\partial B^{2}} Ln Z.
\end{equation}

Mean value of all operators can be related to partition function i. e. $<A>$

\begin{equation}
<A>=\frac{\sum_{i<j<k,\cdots} A e^{-\beta (\in_{i}+\in_{j}+\in_{k}+\cdots )}}{Z}.
\end{equation}

On the other hand, the phenomenological Ising-like model was proposed because of our previous results using classical localized spins 
lead basically to an Ising-like model \cite{VLNA2009,SVNA2009}. High temperature $\chi$ will be compared with experimental results of the 
nickelate one-dimensional compound $Y_{2-n}Ca_{n}BaNiO_{5}$ \cite{DiTusaKojima}. 

\section{Results and discussions}

\begin{figure}[h]
\centering \includegraphics[scale=0.4]{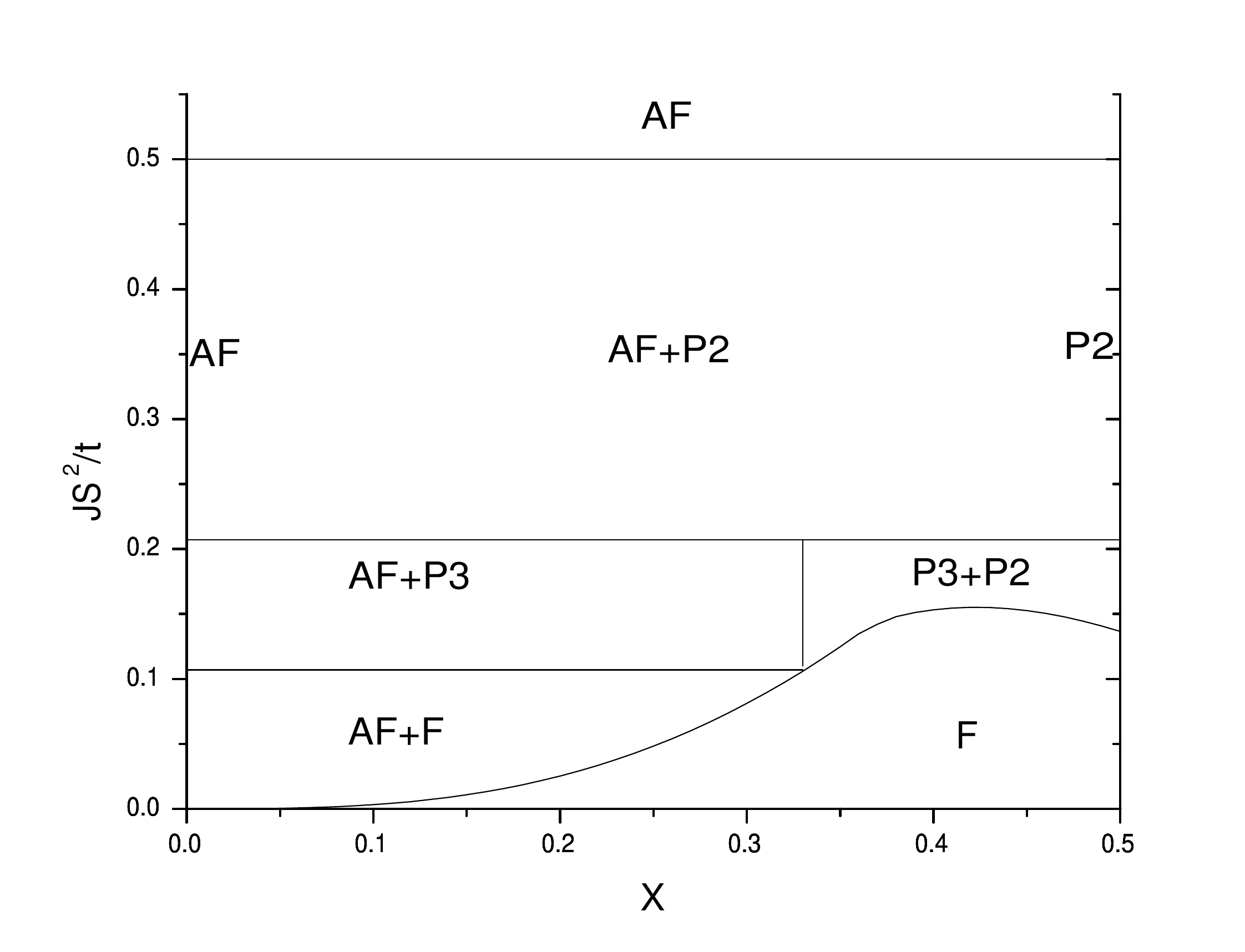}
\caption{Itinerant electron density $x$ vs super-exchange interaction energy $JS^{2}/t$ phase diagram.}
\label{Fig1CL}
\end{figure}

In this section, phase diagram, parallel static magnetic susceptibility MS and short-range spin-spin correlations are presented for a particular 
open linear chain of $N=20$ sites. In the thermodynamic limit, phase diagram is shown in Figure \ref{Fig1CL}. This phase diagram is similar to our 
previous one using classical localized spins (s=3) \cite{VLNA2009,SVNA2009}. Phase separation between ferromagnetic (F) 
$\cdot\cdot\cdot\uparrow\uparrow\uparrow\uparrow\uparrow\uparrow\cdot\cdot\cdot$ and anti-ferromagnetic (AF) 
$\cdot\cdot\cdot\uparrow\downarrow\uparrow\downarrow\uparrow\downarrow\cdot\cdot\cdot$ phases is
found for low super-exchange interaction energy. On the other hand phase separation between P2 
$\cdot\cdot\cdot\uparrow\uparrow\downarrow\downarrow\uparrow\uparrow\cdot\cdot\cdot$ and P3 
$\cdot\cdot\cdot\uparrow\uparrow\uparrow\downarrow\downarrow\downarrow\uparrow\uparrow\uparrow\cdot\cdot\cdot$ phases and the AF phase 
was obtained for high $JS^{2}/t$. Because of the scalar $s=1$ spin character used in this paper canted CP3, CP2 and 
T phases are not obtained in this paper \cite{VLNA2009,SVNA2009}. 
The AF phase observed at $x=0$ was previously studied for an Ising (s=1) and classical (s=3) model respectively in references \cite{F1964}. 
 
\begin{figure}[h]
\centering  \includegraphics[scale=0.8]{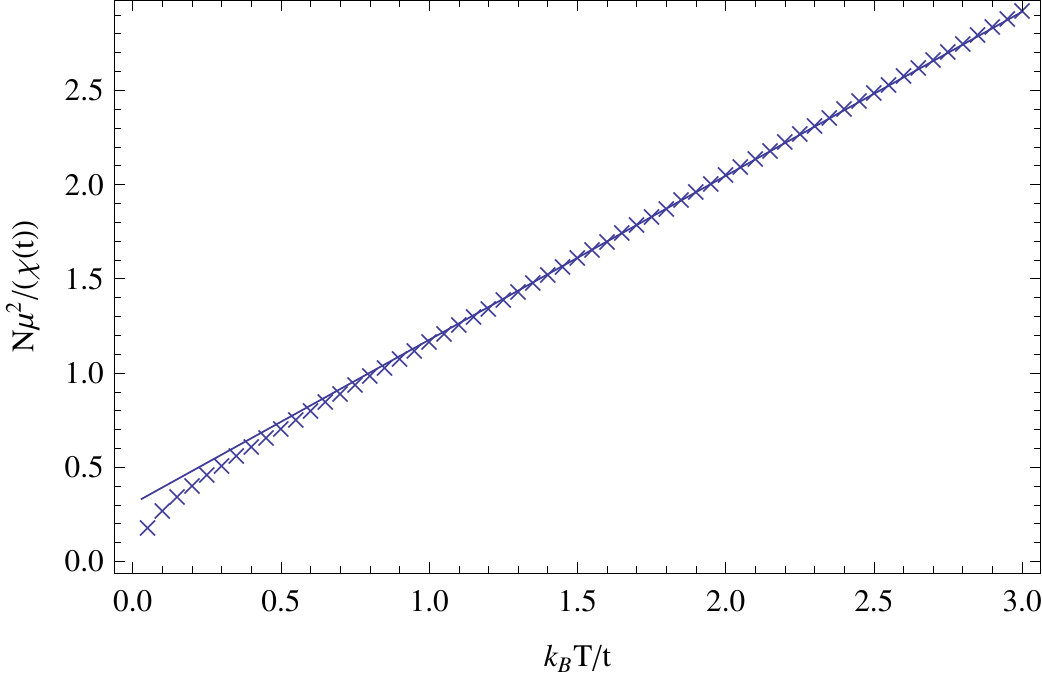}
\caption{Inverse of the magnetic susceptibility $(\chi )$ vs temperature $(k_{B}T/t)$ for $x=0.05$ i.e. one itinerant electron and a 
typical value of the super-exchange interaction energy $JS^{2}/t=0.2$. 
Curie-Weiss like behavior at high temperature limit can be observed. Solid line represents $J_{H} \gg k_{B}T \gg t \gg JS^{2}$ limit.}
\label{Fig2CL}
\end{figure}
\begin{figure}[h]
\centering  \includegraphics[scale=0.8]{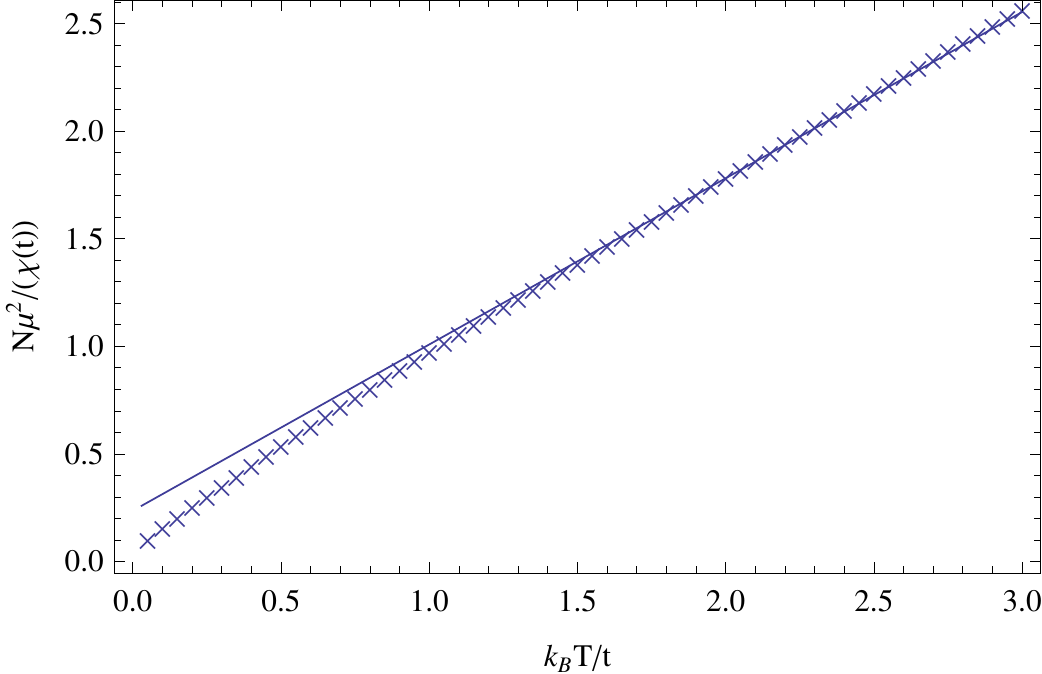}
\caption{The same as Figure \ref{Fig2CL} but for two itinerant electrons $x=0.10$.}
\label{Fig3CL}
\end{figure}
\begin{figure}[h]
\centering  \includegraphics[scale=0.8]{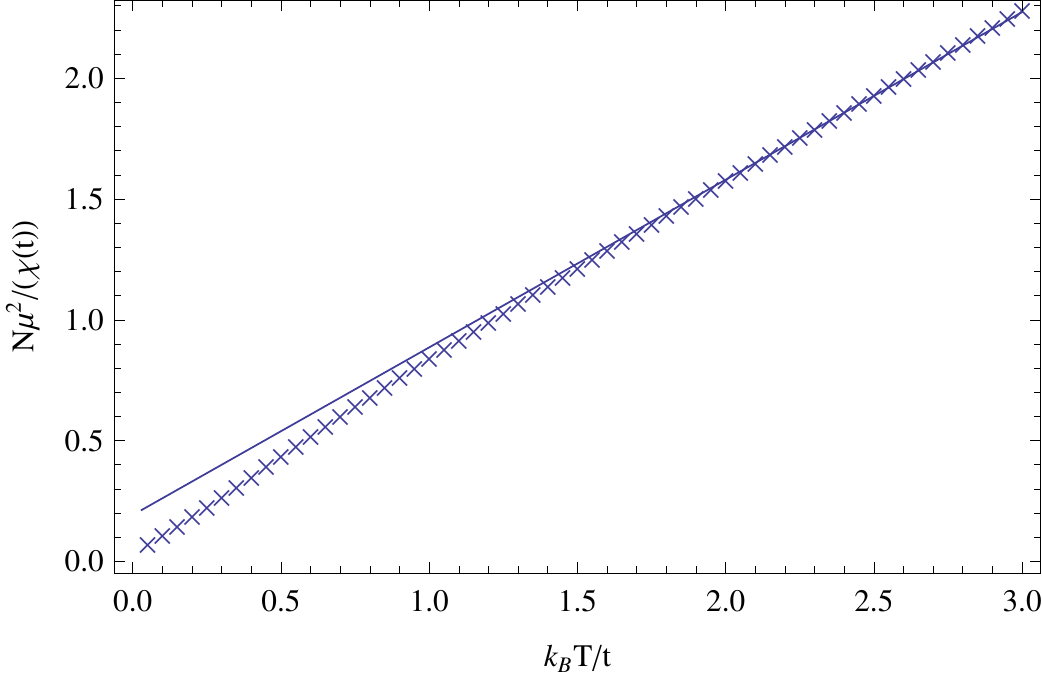}
\caption{The same as Figure \ref{Fig2CL} but for three itinerant electrons $x=0.15$.}
\label{Fig4CL}
\end{figure} 
Figures \ref{Fig2CL}, \ref{Fig3CL} and \ref{Fig4CL} show the inverse of the magnetic susceptibility vs temperature for
one, two and three itinerant electrons respectively. Solid lines in those figures represent high temperature $J_{H} \gg k_{B}T \gg t \gg JS^{2}$ limit.
Curie-Weiss behavior can be easily observed in those figures as 
$\frac{\chi t}{N\mu ^{2}}=\frac{C}{\frac{k_{B}T}{t}+\frac{k_{B}T_{c} }{t}};$ being $C$ Curie constant and $T_{c}$ Curie-Weiss like temperature.
($C=1.15;$ $\frac{k_{B}T_{c} }{t}=0.35$), ($C=1.30;$ $\frac{k_{B}T_{c} }{t}=0.31$) and ($C=1.44;$ $\frac{k_{B}T_{c} }{t}=0.28$) for one, two and three
itinerant electrons respectively. Curie constant can be rigorously extracted for the former limit $J_{H}\rightarrow
\infty $ and $t=J=0$. For this goal it is considered $N$ localized spins and $N_{e}$ itinerant electrons. Because of $J_{H}\rightarrow
\infty $ limit Hilbert space is reduced. So there are $N_{e}$ and $N-N_{e}$ free particles with $\pm 2\mu B$ and $\pm \mu B$ 
energies respectively. Being $B$ the magnetic field. The former gives $\frac{\chi t}{N\mu^{2}}=\frac{1+3x}{\frac{k_{B}T}{t}}$. Curie constant is identified like $1+3x$. It gives
$1.15$, $1.30$ and $1.45$ for one, two and three itinerant electrons respectively i. e. $(x=0.05, x=0.10$ and $x=0.15)$. 
These values are very close to those obtained in figures \ref{Fig2CL}-\ref{Fig4CL}. Now, we can use our Curie constant 1+3x to make contact with results
of the nickelate one-dimensional compound $Y_{2-n}Ca_{n}BaNiO_{5}$. $3n$ (S=1/2) for Curie constant was proposed by Kojima \textit{et al.} \cite{DiTusaKojima}
Kojima \textit{et al.} proposed that each Ca-atom introduces three $S=1/2$ spins. They studied hole dopings $n=0.045, 0.095$ and $0.149$. In our case
these itinerant holes correspond to $x=0.955, 0.905$ and $0.851$ itinerant electrons studied here. 
It means Curie constant (1+3x) as $C=3.865, 3.715$ and $3.553$ or simply $C=4-3n$ if we introduce holes as Kojima. 
$n=0$ corresponds to $C=4$ or Ne=N electrons coupled with N localized
spins S=1/2 by an infinite Hund's coupling. On the other hand, $n=1$ is exactly N localized spins S=1/2 with $C=1$. So, the effect to introduce
holes in our itinerant electron system is to reduce Curie constant. 
For low temperature the model proposed by Kojima \textit{et al.} is very close to our P3+AF phase separation.  
On the other hand, Curie-Weiss like temperature $T_{c}$ decreases as
itinerant electron density increases. Itinerant electrons are responsible for the former F behavior because of our DE interaction.   

\begin{figure}[h]
\centering  \includegraphics[scale=0.8]{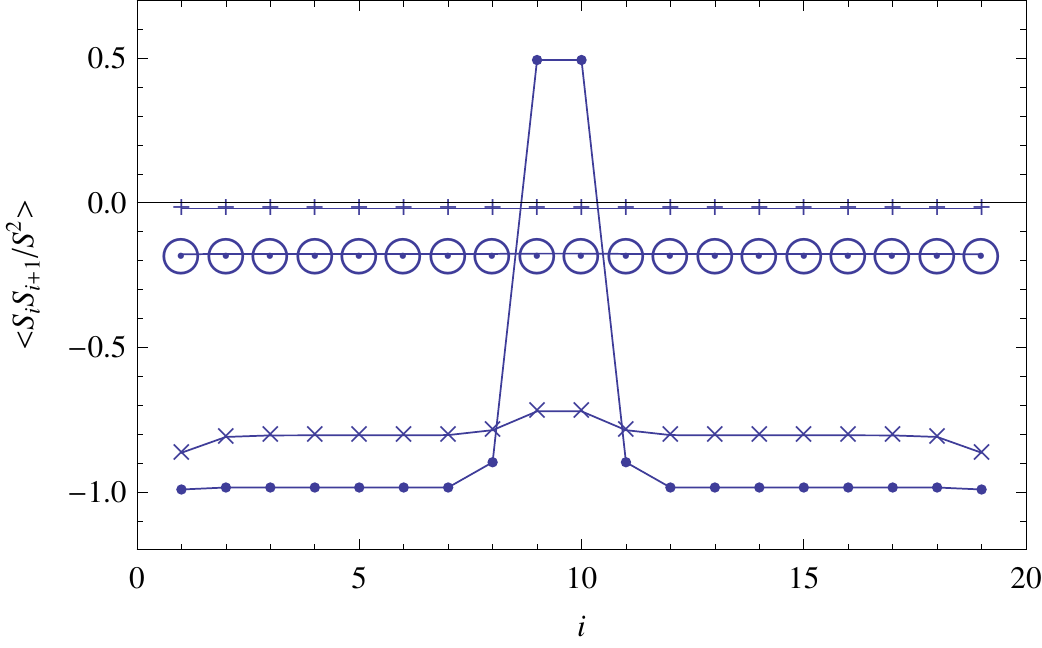}
\caption{Short-range spin-spin correlations within our Ising-like model for one itinerant electron $x=0.05$ and $JS^{2}/t=0.2$.
Solid circles, cross, large open circles and plus symbols respresent four different temperatures $k_{B}T/t=0.01, 0.1, 1.0$ and $10$ respectively.}
\label{Fig5CL}
\end{figure}

\begin{figure}[h]
\centering  \includegraphics[scale=0.8]{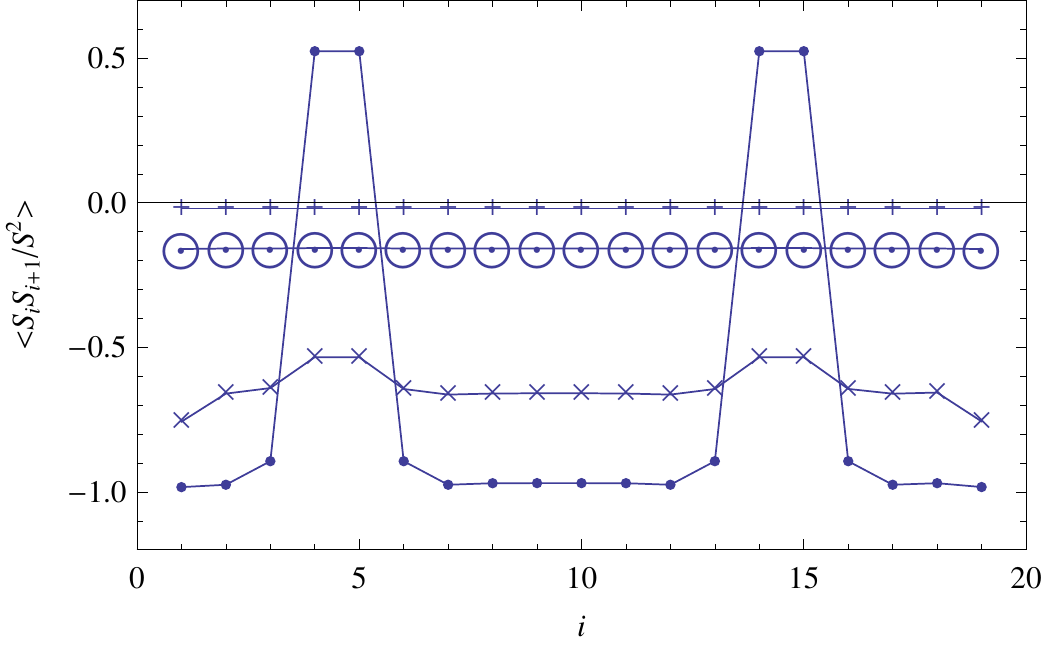}
\caption{The same as figure \ref{Fig5CL} but for two itinerant electrons $x=0.10$.}
\label{Fig6CL}
\end{figure}

\begin{figure}[h]
\centering  \includegraphics[scale=0.8]{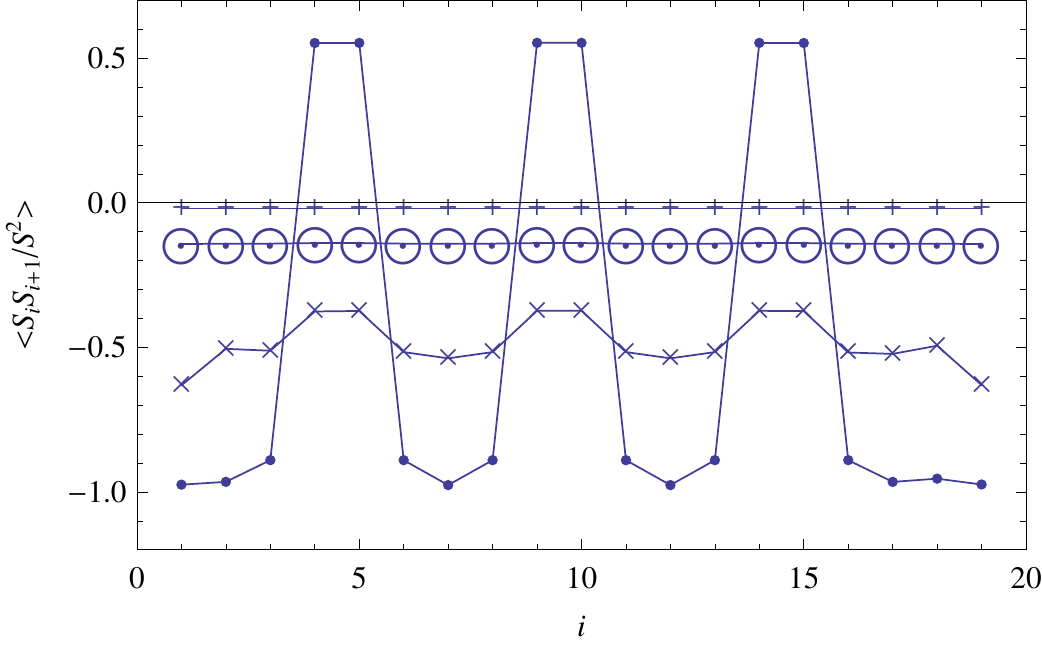}
\caption{The same as figure \ref{Fig5CL} but for three itinerant electrons $x=0.15$.}
\label{Fig7CL}
\end{figure}

Short range spin-spin correlations $<S_{i}S_{i+1}/S^{2}>$ at zero magnetic field can be observed in figures \ref{Fig5CL}-\ref{Fig7CL} 
for a typical value of $JS^{2}/t=0.2$ and four different temperatures $k_{B}T/t=0.01, 0.1, 1.0, 10$ 
solid circles, cross, large open circles and plus symbols respectively were used. To obtain these short range correlations negative in-site 
$(\epsilon/t=-0.1)$ energies were used to pin one, two and three polarons in the linear chain as can be observed in figures \ref{Fig5CL}-\ref{Fig7CL}
respectively. These negative in-site energies can be related with impurities in our linear chain. For low temperature can be clearly seen polarons of three sites in an AF background. Similar polarons were found in 
reference \cite{neuber2006} by using quantum S=3/2 core spins. 
This phase with disordered polarons is degenerated to our P3+AF phase separation. It means ordered polarons of three sites in an AF background. 
At high temperature $(\frac{k_{B}T}{t}>0.1)$ polarons disperse and a very low correlation is observed.
 

\begin{figure}[h]
\centering  \includegraphics[scale=0.8]{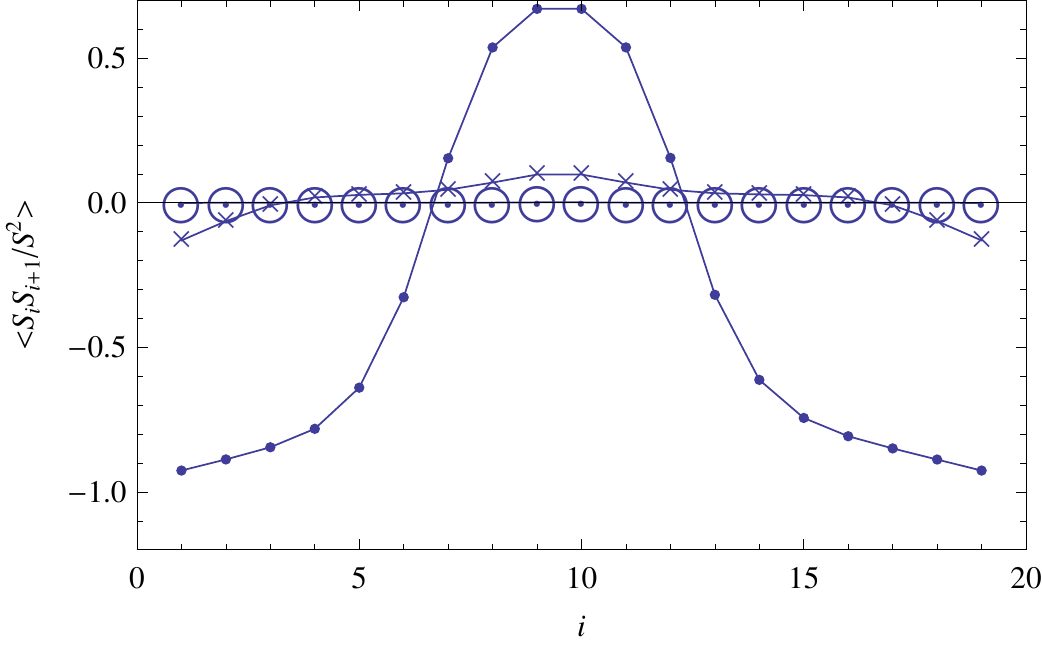}
\caption{Short-range spin-spin correlations for one itinerant electron $x=0.05$ and $JS^{2}/t=0.02$.
Solid circles, cross and large open circles respresent three different temperatures $k_{B}T/t=0.01, 0.1$ and $1.0$ respectively.}
\label{Fig8CL}
\end{figure}

\begin{figure}[h]
\centering  \includegraphics[scale=0.8]{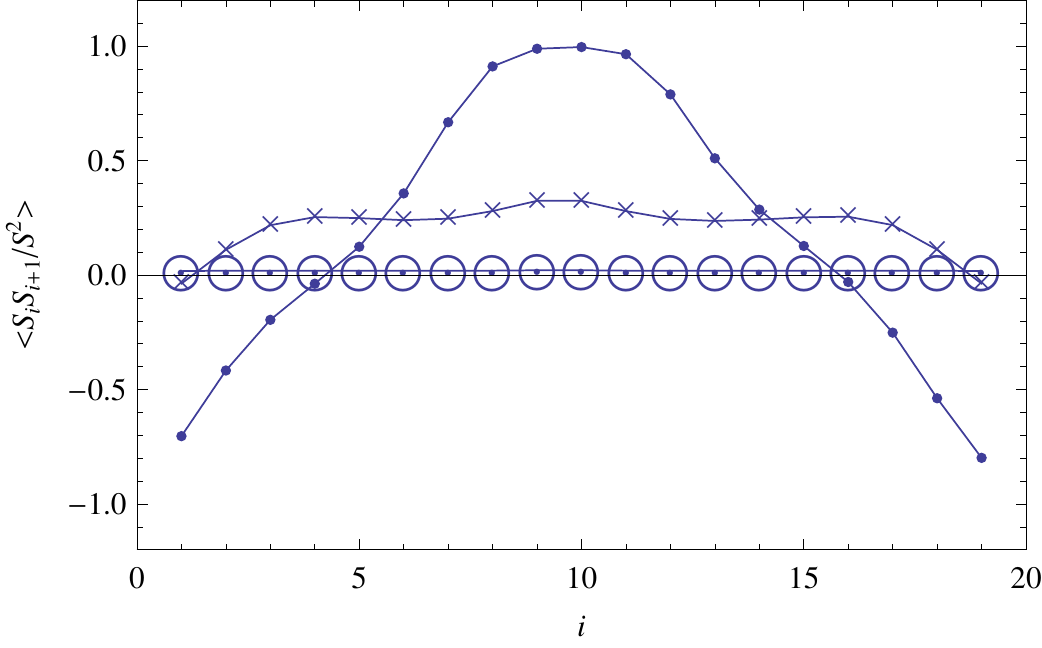}
\caption{The same as figure \ref{Fig8CL} but for two itinerant electrons $x=0.10$.}
\label{Fig9CL}
\end{figure}

\begin{figure}[h]
\centering  \includegraphics[scale=0.8]{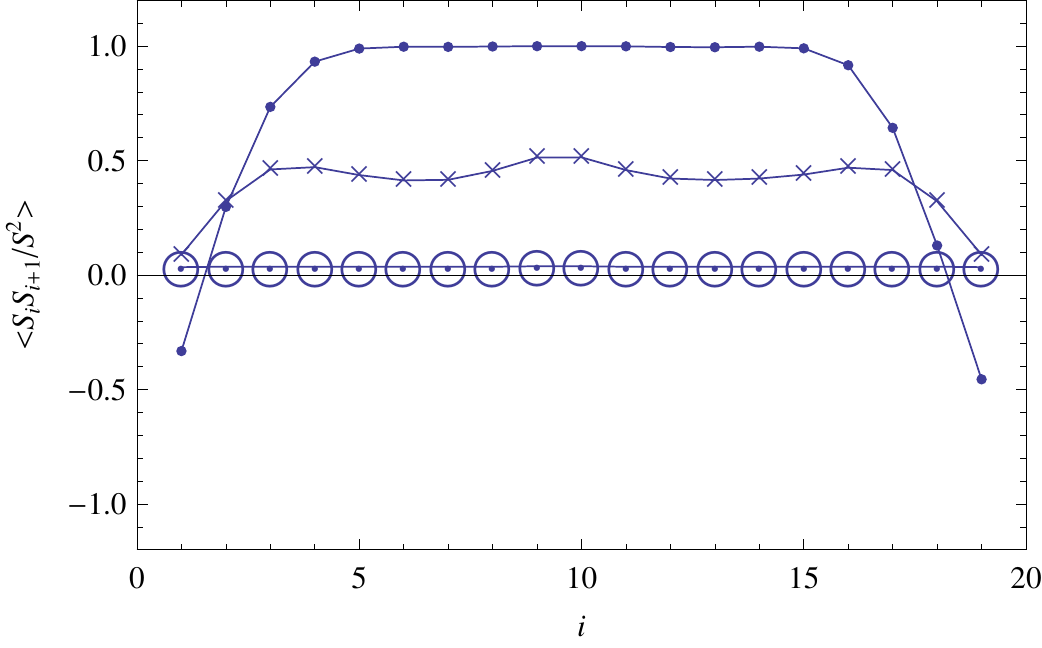}
\caption{The same as figure \ref{Fig8CL} but for three itinerant electrons $x=0.15$.}
\label{Fig10CL}
\end{figure}

In the same way, figures \ref{Fig8CL}-\ref{Fig10CL} show short range spin-spin correlations $<S_{i}S_{i+1}/S^{2}>$
for another typical value of $JS^{2}/t=0.02$ and three different temperatures $k_{B}T/t=0.01, 0.1$ and $1.0$. In this case
only one in-site $(\epsilon/t=-0.1)$ energy was utilized to pin de F phase as can be seen in figures \ref{Fig8CL}-\ref{Fig10CL}.
For low temperature F-AF phase separation can be observed. The F phase increases as the itinerant electron density $x$ increases, see 
figures \ref{Fig8CL}-\ref{Fig10CL}. The former is because of DE interaction. At high temperature the F phase disperses and a very low correlation is observed.
 
\begin{figure}[h]
\centering  \includegraphics[scale=0.8]{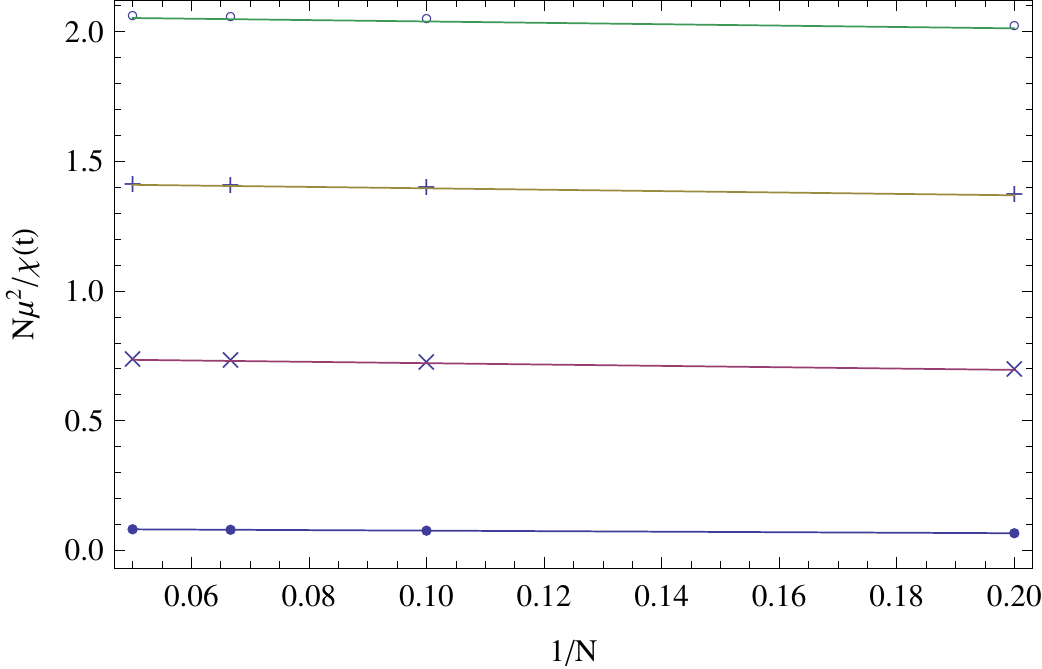}
\caption{Inverse of magnetic susceptibility vs inverse of N sites for an itinerant electron density of $x=0.2$ and $JS^{2}/t=0.2$.
Solid circles, cross, plus and open circles respresent four different temperatures $k_{B}T/t=0.1, 1, 2$ and $3$ respectively. Fitting solid
lines are also shown in the same figure.}
\label{Fig11CL}
\end{figure}
\begin{figure}[h]
\centering  \includegraphics[scale=0.8]{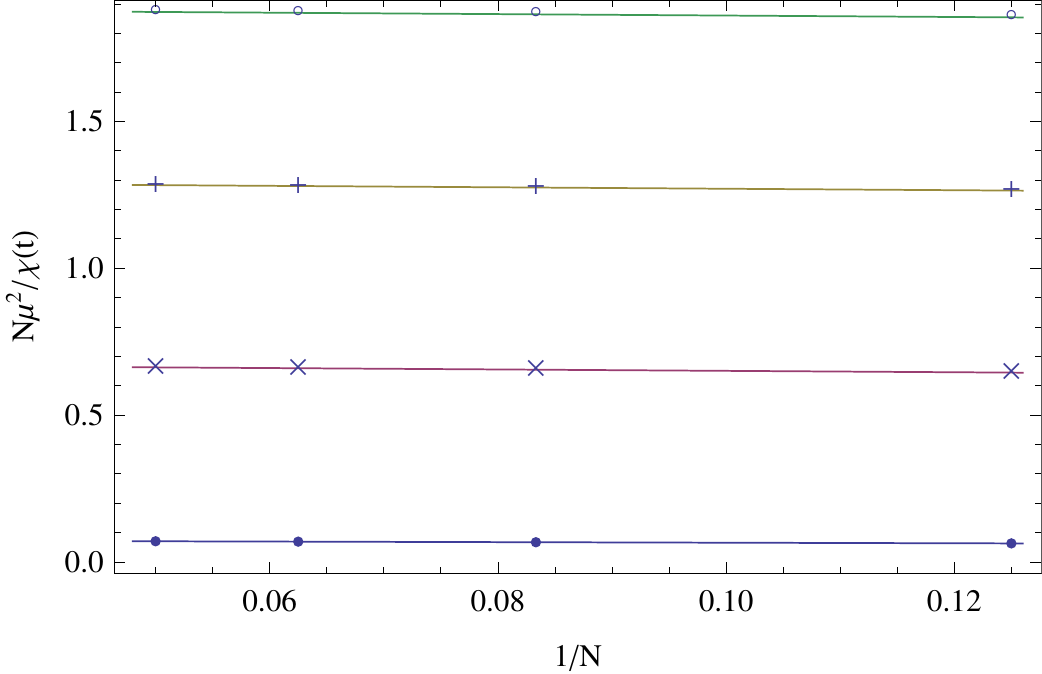}
\caption{The same as figure \ref{Fig11CL} but for an itinerant electron density of $x=0.25$.}
\label{Fig12CL}
\end{figure}
\begin{figure}[h]
\centering  \includegraphics[scale=0.8]{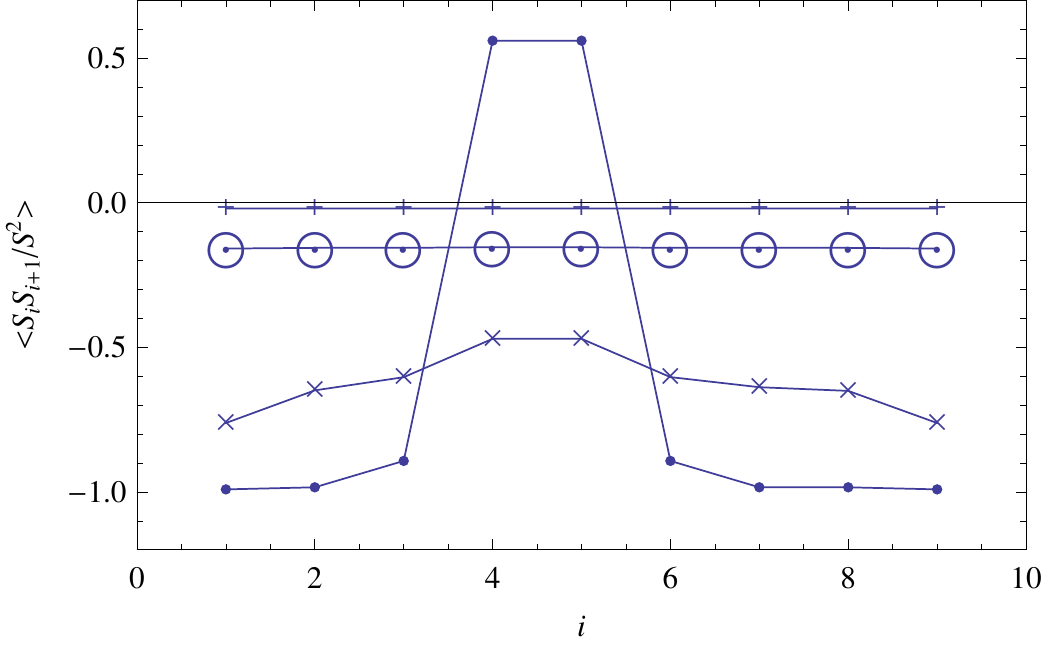}
\caption{The same as figure \ref{Fig6CL} but for a chain of $N=10$ sites, one itinerant electron $x=0.10$ is presented.}
\label{Fig13CL}
\end{figure} 
It is tempting to apply our results to the magnetic
properties of the hole doped $Y_{2-n}Ca_{n}BaNiO_{5}$. Doing so raises the
question of the relation between quantum spins and classical spins cases. It
is clear that some properties are specific to the quantum character of the
spins, in particular the Haldane gap occurring in Heisenberg $S=1$ chains,
as in the case of un-doped $Y_{2}BaNiO_{5}$. However, in the doped case the
itineracy of doped electrons or holes plays an important role taken into
account by the double-exchange mechanism. The essential behavior of the spin
correlations in the quantum level is similar in the classical case. For the
commensurate filling $x=1/2$ the polaronic phase $P2$ reference \cite{VLNA2009} is qualitatively similar to the
quantum $S=1/2$ case.

We have calculated magnetic susceptibility for typical values of the conduction electron density
to make contact with experiments\cite{DiTusaKojima}. The inverse of magnetic susceptibility (%
$\chi $) vs $T$ presents a complicated behavior as described in the former lines. 
At high temperature Curie-Weiss behavior was obtained. As shown, Curie constant is basically t-J independent. 
Our Ising-like results give $C=1+3x$ or $C=4-3n$. 
Kojima \textit{et al.} from experimental results 
proposed $C\simeq 3n$ (S=1/2). In our case we remove electrons from an S=1 system $n=0$. In the case of Kojima, holes are added.
In this case our Ising-like model may be can be related with experimental results. 
Curie-Weiss temperature $T_{c}$ is t-J dependent and can be related with Curie-like behavior observed in this compound \cite{DiTusaKojima}. 
It is important to mention that the contribution related to the
Haldane gap in $S=1$ spin chains decreases exponentially with decreasing
temperature and becomes negligible at low temperature $T<20K$ \cite{Das2004}. It is
difficult to identify the different contributions to the magnetic
susceptibility in such a complex magnetic ground state. Of course, our
comparison with the experimental results becomes irrelevant below the
spin-glass transition identified to be $T_{g}\sim 2.9K$. Finite size effects are taken into account to show that our $N=20$ sites
are of relevance. Inverse of magnetic susceptibility vs inverse of N sites for different temperatures
are shown in figures \ref{Fig11CL} and \ref{Fig12CL} for an itinerant electron density of $x=0.2$ and $x=0.25$ respectively. 
Fitting solid lines $\alpha+\beta(1/N)$ with an error of $10^{-4}$, 95 per cent of confidence levels are shown in the same figures. 
As can be seen in the same figures an error of $\beta(1/N)\sim 10^{-2}$ is obtained if $N=20$ sites are taken into account. 
The $t=J=0$ limit, that is N-site independent, is also compared with these thermodynamic limits, giving an error of $10^{-1}$ \cite{Aclar}. 
Finite size effects for a Heisenberg and an Ising model (without itinerant electrons $x=0$) were studied in reference \cite{F1964}.
As can be seen in that reference, magnetic susceptibility is almost N-site independent at high temperature limit. In our model, 
because of itinerant electrons, both high and low temperature limits lead to the same qualitative behavior.
 
It is also presented, in figure \ref{Fig13CL}, short-range spin-spin correlations for one itinerant electron
and $N=10$ sites $x=0.1$ and $JS^{2}/t=0.2$. These results can be compared with results shown in figure \ref{Fig6CL} for $N=20$ sites
and two itinerant electrons. The same spin-spin correlations behavior can be observed. Magnetic phase diagram for classical localized spins
and an exchange model, as used in this paper, is compared with the thermodynamic limit in reference \cite{VLNA2009}. As can be observed 
in that reference, the same magnetic phases were obtained. 
   
Of course that because of our exact results very long systems cannot be studied easily because of a huge CPU time used.

\section{Conclusions}

In this work, we presented exact parallel static magnetic susceptibility calculations and short-range spin-spin correlations 
of an equivalent Ising-like DE+SE model using large Hund's coupling. 
Magnetic susceptibility was calculated in a region where P3-AF and F-AF phase separation can be found.  
At high temperature Curie-Weiss behavior and a very low correlated system were obtained.  
Curie constant is basically t-J independent and could be related with the Curie-like behavior observed in the nickelate one-dimensional 
compound $Y_{2-n}Ca_{n}BaNiO_{5}$. Finite size effects were considered to show the relevance of our finite $N=20$ system.  \newline
\bigskip

\begin{center}
\textbf{ACKNOWLEDGMENT}
\end{center}

We want to acknowledge partial support from CONACyT Grant-57929 and
PAPIIT-IN108907 from UNAM. E.V wants to acknowledge Dr. Calderon for using the cluster.  


\begin{thebibliography}{99}
\bibitem{Aclar}
It is important to mention that another itinerant electronic densities and super-exchange couplings were considered. 
Almost the same errors were obtained and the same qualitative behavior in magnetic susceptibility and spin-spin correlations.
\end{thebibliography}
\end{document}